\documentclass[pdftex,twocolumn,epjc3]{svjour3}          

\RequirePackage[T1]{fontenc}

\smartqed  
\RequirePackage{graphicx}
\RequirePackage{mathptmx}      
\RequirePackage{flushend}
\RequirePackage[numbers,sort&compress]{natbib}
\RequirePackage[colorlinks,citecolor=blue,urlcolor=blue,linkcolor=blue]{hyperref}
\RequirePackage{epstopdf}
\journalname{Eur. Phys. J. C}
\RequirePackage{amsmath,amssymb}
\begin{document}

\title{ Gauge field and brane-localized kinetic terms on the chiral square}


\author{Ricardo G. Landim\thanksref{e1,addr1}
}

\thankstext{e1}{ricardo.landim@tum.de}

\institute{Physik Department T70, James-Franck-Strasse,
Technische Universit\"at M\"unchen, 85748 Garching, Germany\label{addr1}
}

\date{Received: date / Accepted: date}

\maketitle

\begin{abstract}
Extra dimensions have been used as attempts to explain several phenomena in particle physics. In this paper we investigate the role of brane-localized kinetic terms  (BLKT) on thin and thick branes with two flat extra dimensions (ED) compactified on the chiral square, and an abelian gauge field in the bulk. The results for a thin brane have resemblance with the 5-D case, leading to a tower of massive KK particles whose masses depend upon the compactification radius and the BLKT parameter. On the other hand, for the thick brane scenario, there is no solution that satisfy the boundary conditions. Because of this, the mechanism of suppressed couplings due to ED \cite{Landim:2019epv} cannot be extended to 6-D. \end{abstract}

\section{\label{sec:intro}Introduction}

Extra dimensions (ED) have been considered over the decades as tools to address a wide range of issues in particle physics, such as   the  hierarchy 
\cite{Antoniadis:1990ew,Dienes:1998vh,Antoniadis:1998ig,ArkaniHamed:1998rs,Randall:1999ee,Arkani-Hamed:2016rle} and flavor problems 
\cite{Agashe:2004cp,Huber:2003tu,Fitzpatrick:2007sa}. Quantum field theory with two ED, for instance, may provide explanations for proton stability \cite{Appelquist:2001mj},  origin of electroweak symmetry breaking \cite{ArkaniHamed:2000hv,Hashimoto:2000uk,Csaki:2002ur,Scrucca:2003ut}, breaking of grand unified gauge groups \cite{Hebecker:2001jb,Hall:2001xr,Asaka:2002my,Asaka:2003iy} and  the number of fermion
generations \cite{Dobrescu:2001ae,Fabbrichesi:2001fx,Borghini:2001sa,Fabbrichesi:2002am,Frere:2001ug,Watari:2002tf}. The Standard Model (SM) itself might be extended   by employing ED, in the so-called Universal Extra Dimension model (UED). In this context, the  whole SM content is
promoted to fields which propagate in compact ED, having Kaluza-Klein (KK) excitations, in either one \cite{Appelquist:2000nn} or two ED \cite{Dobrescu:2004zi,Burdman:2005sr,Ponton:2005kx,Burdman:2006gy}. The zero-mode of each KK tower of states in 4-D is thus identified with the 
correspondent SM particle and a lowest KK state can be a dark matter (DM) candidate. Current results from LHC \cite{Aad:2015mia,TheATLAScollaboration:2013uha} impose bounds on the UED compactification radius $R$ for one  ($R^{-1}>1.4-1.5$ TeV)\footnote{For $\Lambda R \sim 5-35$, where $\Lambda$ is
the cutoff scale.}
\cite{Deutschmann:2017bth,Beuria:2017jez,Tanabashi:2018oca} or two ED ($R^{-1}>900$ GeV) \cite{Burdman:2016njl}.  

On the other hand, the 4-D gravity  might be an emergent phenomenon from ED, as in the DGP model \cite{Dvali:2000hr}, where the brane-induced term  was initially obtained for a massless spin-2 field. Such a  mechanism is possible for a spin-1 field as well, in which a brane-localized kinetic term (BLKT) is generated on the brane by radiative corrections  due to the interaction of localized matter fields on the brane with the gauge field in the bulk \cite{Dvali:2000rx}, and it holds for infinite-volume, warped and compact ED. The  same mechanism also works  for two ED  \cite{Dvali:2000xg,Dvali:2001ae,Dvali:2002pe} and the role of such a term has been investigated in several different scenarios \cite{Carena:2002me,Carena:2002dz,delAguila:2003gv,delAguila:2003bh,Davoudiasl:2002ua,Davoudiasl:2003zt, Rizzo:2018ntg,Rizzo:2018joy}. The localization of matter/gauge fields in branes was studied in other contexts, for thin \cite{ArkaniHamed:1998rs,Dvali:1998pa, Alencar:2014moa,Alencar:2017dqb,Alencar:2015awa,Alencar:2015rtc,Alencar:2015oka,Alencar:2018cbk,Freitas:2018iil} and thick branes  \cite{DeRujula:2000he,Georgi:2000wb}.

ED can also be employed in order to elucidate the nature of the DM and its possible interaction with the SM. Usually, a DM candidate may couple with the SM through a scalar mediator (or directly through Higgs if DM is a scalar field), via the so-called Higgs portal \cite{Silveira:1985rk,McDonald:1993ex,Burgess:2000yq,Bento:2000ah,Bertolami:2007wb,Bento:2001yk,MarchRussell:2008yu,Biswas:2011td,Costa:2014qga,Eichhorn:2014qka,Khan:2014kba,Queiroz:2014yna,Kouvaris:2014uoa,Bhattacharya:2016qsg,Bertolami:2016ywc,Campbell:2016zbp,Heikinheimo:2016yds,Kainulainen:2016vzv,Nurmi:2015ema,Tenkanen:2016twd,Casas:2017jjg,Cosme:2017cxk,Heikinheimo:2017ofk,Landim:2017kyz}, or through a vector mediator, which is introduced by a kinetic-mixing term \cite{Holdom:1985ag,Holdom:1986eq,Dienes:1996zr,DelAguila:1993px,Babu:1996vt,Rizzo:1998ut,Feldman:2006wd,Feldman:2007wj,Pospelov:2007mp,Pospelov:2008zw,Davoudiasl:2012qa,Davoudiasl:2012ag,Essig:2013lka,Izaguirre:2015yja}. In both cases, much of the parameter space has been excluded by a diverse set of experiments and observations \cite{Duerr:2015aka,Athron:2017kgt,Djouadi:2011aa,Cheung:2012xb,Djouadi:2012zc,Cline:2013gha,Endo:2014cca,Goudelis:2009zz,Urbano:2014hda,Akerib:2015rjg,He:2016mls,Escudero:2016gzx,Ade:2015xua,Cline:2013fm,Slatyer:2015jla,Ackermann:2015zua,Akerib:2016vxi,Tan:2016zwf,Agnese:2014aze,Aprile:2012nq,Aartsen:2012kia,Aartsen:2016exj,Battaglieri:2017aum,Riordan:1987aw,Bjorken:1988as,Bross:1989mp,Bjorken:2009mm, Pospelov:2008zw,Davoudiasl:2012ig,Endo:2012hp,Babusci:2012cr,Archilli:2011zc,Adlarson:2013eza,Abrahamyan:2011gv,Merkel:2011ze,Reece:2009un,Aubert:2009cp,Dreiner:2013mua}. The small value of both couplings constants may be explained if we consider a single, flat ED and a thick brane with BLKT spread inside it \cite{Landim:2019epv}, where inside the `fat' brane the SM fields behaves as in the UED model with one ED. 

An obvious generalization of this previous work then would be investigate the possibility of suppressed couplings, along with the presence of BLKT on thin branes, for higher dimensional spacetimes. In this paper we investigate this possibility in 6-D, which is a natural extension since UED has been built for two ED as well. Alongside with this aim, we consider BLKT on thin branes, leading to results  that can be compared with the 5-D case \cite{Carena:2002me}. We assume the same compactification of the UED model in 6-D, where the so-called chiral square was chosen because it is the simplest compactification that leads to chiral quarks and leptons in 4-D \cite{Dobrescu:2004zi}. For simplicity we will only consider an abelian gauge field in the bulk, although for other fields the results are analogue. The presence of BLKT on thin branes has a similar result as in the 5-D case, where the masses of the 4-D KK tower of states are determined by a transcendental equation. A thick brane with a BLKT, on the other hand, is not allowed by the boundary conditions (BC) at the intersection between the regions thick brane/bulk. Therefore, the mechanism in 5-D can be consistently extended for 6-D only for thin branes.

This paper is organized as follows. Sect.  \ref{sec:gauge} reviews the  6-D gauge 
field without any BLKT on the chiral square. In Sect. \ref{blkt-thin}  we introduce thin branes with BLKT and analyze the spectrum of masses while in Sect. \ref{blkt-thick} a fat brane is considered. Sect. \ref{sec:discussion} is reserved for conclusions.

\section{Gauge field in the bulk}\label{sec:gauge}

We will consider two flat and transverse ED ($x^4$ and $x^5$) compactified on the chiral square. The square has size $\pi R$ and the adjacent sides are identified $(0,y)\sim (y,0)$ and $(\pi R,y)\sim (y,\pi R)$, with $y\in [0,\pi R]$, which means the Lagrangians at those points have the same values for any field configuration: $\mathcal{L}(x^\mu, 0,y)=\mathcal{L}(x^\mu,y,0)$ and $\mathcal{L}(x^\mu, \pi R,y)=\mathcal{L}(x^\mu,y,\pi R)$.

There is only an abelian gauge field $V^A, ~A=0-3, 4, 5$ in the bulk  and the action is similar to the one of UED with two ED \cite{Dobrescu:2004zi,Burdman:2005sr}, given by
\begin{equation}\label{eq:actionV}
    S=\int d^4x\int_0^{\pi R} dx^4\int_0^{\pi R}dx^5 \left( -\frac{1}{4}V_{AB}V^{AB}+\mathcal{L}_{GF} \right)\,,
\end{equation}
where A is the 6-D index and the gauge fixing term has the following form to cancel the mixing between $V_4$ and $V_5$ with $V_\mu$ \cite{Burdman:2005sr}
\begin{equation}\label{eq:actionGF}
\mathcal{L}_{GF}=-\frac{1}{2\xi}\Big[\partial_\mu V^\mu-\xi(\partial_4V_4+\partial_5V_5)\Big]^2\,,
\end{equation}
where  $\xi$ is the gauge fixing parameter and we will work in the Feynman gauge ($\xi=1$). After integrating by parts the action (\ref{eq:actionV}) is written as
\begin{eqnarray}\label{eq:actionV2}
    S=&\int d^4x\int_0^{\pi R} dx^4\int_0^{\pi R}dx^5 \bigg\{ -\frac{1}{4}V_{\mu\nu}V^{\mu\nu}-\frac{1}{2\xi}(\partial_\mu V^\mu)^2\nonumber\\& +\frac{1}{2}\Big[(\partial_4 V_\mu)^2+(\partial_5 V_\mu)^2\Big]+\frac{1}{2}\Big[(\partial_\mu V_4)^2+(\partial_\mu V_5)^2\nonumber\\&-\xi(\partial_4V_4+\partial_5V_5^2)-(\partial_4V_5-\partial_5V_4)^2\Big]\\\nonumber&+\text{surface terms}\bigg\}\,.
\end{eqnarray}

In the Feynman gauge, the equations of motion for the components of $V^A$ are
\begin{equation}
    \label{eq:eqofmotAmu}
   (\Box -\partial_4^2-\partial_5^2)V_A=0\,,
\end{equation}
where $\Box\equiv \partial_\mu\partial^\mu$. Furthermore, it is required that the  surface terms vanish on the boundary, in order not to have flow of energy or momentum across it, \textit{i.e.}
\begin{eqnarray}\label{eq:surfaceterms}
   &\int d^4 x \biggl\{\int dx^4 \Big[V_{5\mu}\delta V^\mu+V_{45} \delta V_4+(\partial_\mu V^\mu-\partial_4V_4\nonumber\\&-\partial_5V_5)\delta V_5 \Big]\biggr\rvert_{x^5=0}^{x^5=\pi R}
  \nonumber\\
    & + \int dx^5 \Big[V_{4\mu}\delta V^\mu-V_{45} \delta V_5+(\partial_\mu V^\mu-\partial_4V_4\nonumber\\
    &-\partial_5V_5)\delta V_4 \Big]\biggr\rvert_{x^4=0}^{x^4=\pi R}\biggr\}=0\,.
\end{eqnarray}
Vanishing the surface terms lead to the following BC for $V_\mu$
\begin{eqnarray}\label{eq:bcAmu}
    V_\mu(y,0)&=V_\mu(0,y)\nonumber\,,\\
    \partial_4 V_\mu |_{(x^4,x^5)=(y,0)}&= \partial_5 V_\mu |_{(x^4,x^5)=(0,y)}\,,\\
     \partial_5 V_\mu |_{(x^4,x^5)=(y,0)}&=- \partial_4 V_\mu |_{(x^4,x^5)=(0,y)}\,,\nonumber
\end{eqnarray}
and for $V_4$ and $V_5$
\begin{eqnarray}\label{eq:bcA4}
    V_4(y,0)&=V_5(0,y)\nonumber\,,\\
       \partial_4 V_4 |_{(x^4,x^5)=(y,0)}&= \partial_5 V_5 |_{(x^4,x^5)=(0,y)}\,,\\
     \partial_5 V_4 |_{(x^4,x^5)=(y,0)}&=- \partial_4 V_5 |_{(x^4,x^5)=(0,y)}\,,\nonumber
\end{eqnarray}
\begin{eqnarray}\label{eq:bcA5}
    V_5(y,0)&=-V_4(0,y)\nonumber\,,\\
       \partial_4 V_5 |_{(x^4,x^5)=(y,0)}&=- \partial_5 V_4 |_{(x^4,x^5)=(0,y)}\,,\\
     \partial_5 V_5 |_{(x^4,x^5)=(y,0)}&= \partial_4 V_4 |_{(x^4,x^5)=(0,y)}\,,\nonumber
\end{eqnarray}
for any $0\leq y \leq \pi R$. The same relations exist for the fields at $(y, \pi R)$ and $(\pi R, y)$. From the above relations it is possible to see the transformation law $(V_4,V_5) \rightarrow (V_5,-V_4)$ satisfied by the fields under $(x^4,x^5)\rightarrow (-x^5,x^4)$ \cite{Burdman:2005sr}.

We expand the components of the 6-D gauge field in KK towers of states
  \begin{equation}\label{eq:KKexpAmu}
      V_\mu(x^\nu, x^4,x^5)=\sum_j\sum_k v_0^{(j,k)}(x^4,x^5) V_\mu^{(j,k)}(x^\nu)\,,
  \end{equation}

  \begin{equation}\label{eq:KKexpA4}
      V_4(x^\nu, x^4,x^5)=\sum_j\sum_k v_4^{(j,k)}(x^4,x^5) V_4^{(j,k)}(x^\nu)\,,
  \end{equation}
  
  \begin{equation}\label{eq:KKexpA5}
      V_5(x^\nu, x^4,x^5)=\sum_j\sum_k v_5^{(j,k)}(x^4,x^5) V_5^{(j,k)}(x^\nu)\,,
  \end{equation}
  which yields the equation of motion for $v_i^{(j,k)}(x^4,x^5)$, with $i=0, 4$ or $5$,
  \begin{equation}\label{eq:vx4x5}
     [\partial_4^2+\partial_5^2+(M_i^{(j,k)})^2]v_i^{(j,k)}(x^4,x^5)=0\,,
  \end{equation}
  where $M_i^{(j,k)}$ are the physical masses of the gauge field $ V_\mu$ and the scalar fields $V_4$ and $V_5$, respectively. The solutions for the equation of motion, which satisfy the BC above, and are normalized through the relation 
  \begin{eqnarray}\label{eq:normalizationcondit}
     \int_0^{\pi R} dx^4\int_0^{\pi R} dx^5\, v_i^{(j,k)}(x^4,x^5)v_i^{(j',k')}(x^4,x^5)&=\delta_{j,j'}\delta_{k,k'}\,,
\end{eqnarray}
  are given by
  \begin{equation}\label{eq:v0}
     v_0^{(j,k)}(x^4,x^5)=\frac{1}{\pi R}\left[\cos (m_j x^4+m_k x^5)\pm\cos(m_k x^4-m_j x^5)\right]\,,
  \end{equation}
    \begin{equation}\label{eq:v4}
     v_4^{(j,k)}(x^4,x^5)=\frac{\sqrt{2}}{\pi R}\sin \Big(\frac{j x^4+k x^5}{R}\Big)\,,
  \end{equation}
      \begin{equation}\label{eq:v5}
     v_5^{(j,k)}(x^4,x^5)=-\frac{\sqrt{2}}{\pi R}\sin \Big(\frac{k x^4-j x^5}{R}\Big)\,,
  \end{equation}
  where $j$ and $k$ are integers and the parameters $m_j$ and $m_k$ are $m_j=j/R$ and $m_k=k/R$, for the $+$ sign in Eq. (\ref{eq:v0}) or $m_j=(j+1/2)/R$ and $m_k=(k+1/2)/R$ for the $-$ sign. 
The physical masses  of the scalar fields are $ (M_{4,5}^{(j,k)})^2=(j^2+k^2)/R^2$ while for the  tower of states of the 4-D vector field they are  given by
\begin{equation}\label{mass}
   (M_0^{(j,k)})^2=m_j^2+m_k^2\,,
\end{equation}
  Unlike the  5-D case, where the new scalar field, which is the extra component of the vector field, can be gauged away, in 6-D there is an additional degree of freedom that remains. This fact is explicitly seen if one works in the unitary gauge, where only one of the two linear combinations of the the scalar fields $V_4$ and $V_5$ is eaten by the vector boson $V_\mu^{(j,k)}$  \cite{Burdman:2005sr}.
  
\section{BLKT on thin branes}\label{blkt-thin}
Applying the same ideas of the last section, we will now consider the effect of BLKT on branes localized at the points $(0,0)$, $(\pi R,\pi R)$ and $(\pi R,0)\sim (0,\pi R)$. We should recall that preserving KK parity implies that operators at $(0,0)$ and  $(\pi R,\pi R)$  are identical. 

\subsection{BLKT at (0,0)
}

We will first analyze the change in the wave-function due to the presence of a BLKT term on a brane localized at $(0,0)$.\footnote{As explained in \cite{Dvali:2001ae} the propagator of the 6-D gauge field is found after a regularization procedure.} The localized kinetic term is four-dimensional for distances shorter than $R$, and it is given by 
\cite{Dvali:2000rx,Dvali:2001ae}
\begin{equation}\label{eq:lagrangianBLKTthin}
\mathcal{L}_{BLKT}=\left[-\frac{1}{4}V_{\mu\nu}V^{\mu\nu}-\frac{1}{2\xi}(\partial_\mu V^\mu)^2\right]\cdot \delta_A R^2\,\delta(x^4,x^5)\,,
\end{equation}
where we conveniently added a gauge-fixing term. After expanding the 6-D gauge field into a tower of KK states, the equation of motion for the wave-function $  v_0^{(j,k)}(x^4,x^5)$ has the same structure of the 5-D case
\begin{equation}\label{eq:vx4x5delta}
     \Big[\partial_4^2+\partial_5^2+M_{j,k}^2+ M_{j,k}^2\delta_AR^2\delta(x_4,x_5)\Big]v_0^{(j,k)}(x^4,x^5)=0\,,
  \end{equation}
  where we relabeled $M_0^{(j,k)}\equiv M_{j,k}$.  
  
  The 4-D Lagrangian is found integrating the wave-function over the ED. The resulting Lagrangian has diagonal terms
\begin{equation}
    \mathcal{L}_4=\sum_{j,k} \biggr [-\frac{1}{4}Z_{(j,k)}V_{\mu\nu}^{(j,k)}V^{\mu\nu}_{(j,k)}
    +Z_{(j,k)}M_{j,k}^2V_\mu^{(j,k)}V^{\mu}_{(j,k)} \biggr]\,,
\end{equation}
where $Z_{(j,k)}$ is a normalization factor, if the wave-function satisfies the relations
 \begin{eqnarray}\label{eq:normalizationconditdelta}
     &\int_0^{\pi R} dx^4\int_0^{\pi R} dx^5\,\Big[1+\delta_AR^2\delta(x^4,x^5)\Big] v_0^{(j,k)}v_0^{(j',k')}=Z_{(j,k)}\delta_{j,j'}\delta_{k,k'}\,,\nonumber\\
     & \int_0^{\pi R} dx^4\int_0^{\pi R} dx^5\,\Big[\partial_4v_0^{(j,k)}\partial_4v_0^{(j',k')}\nonumber\\
    &+\partial_5v_0^{(j,k)}\partial_5v_0^{(j',k')}\Big]=Z_{(j,k)}M_{j,k}^2\delta_{j,j'}\delta_{k,k'} \,.
\end{eqnarray}
The normalization factor for a delta-function at the origin is
\begin{equation}
    Z_{(j,k)}=1+\delta_A R^2 v_0^{(j,k)}(0,0)\,,
\end{equation}
and the gauge field in 4-D becomes canonically normalized after dividing it by $Z_{(j,k)}^{-1/2}$.
   
\begin{figure}
    \centering 
    \includegraphics[scale=0.45]{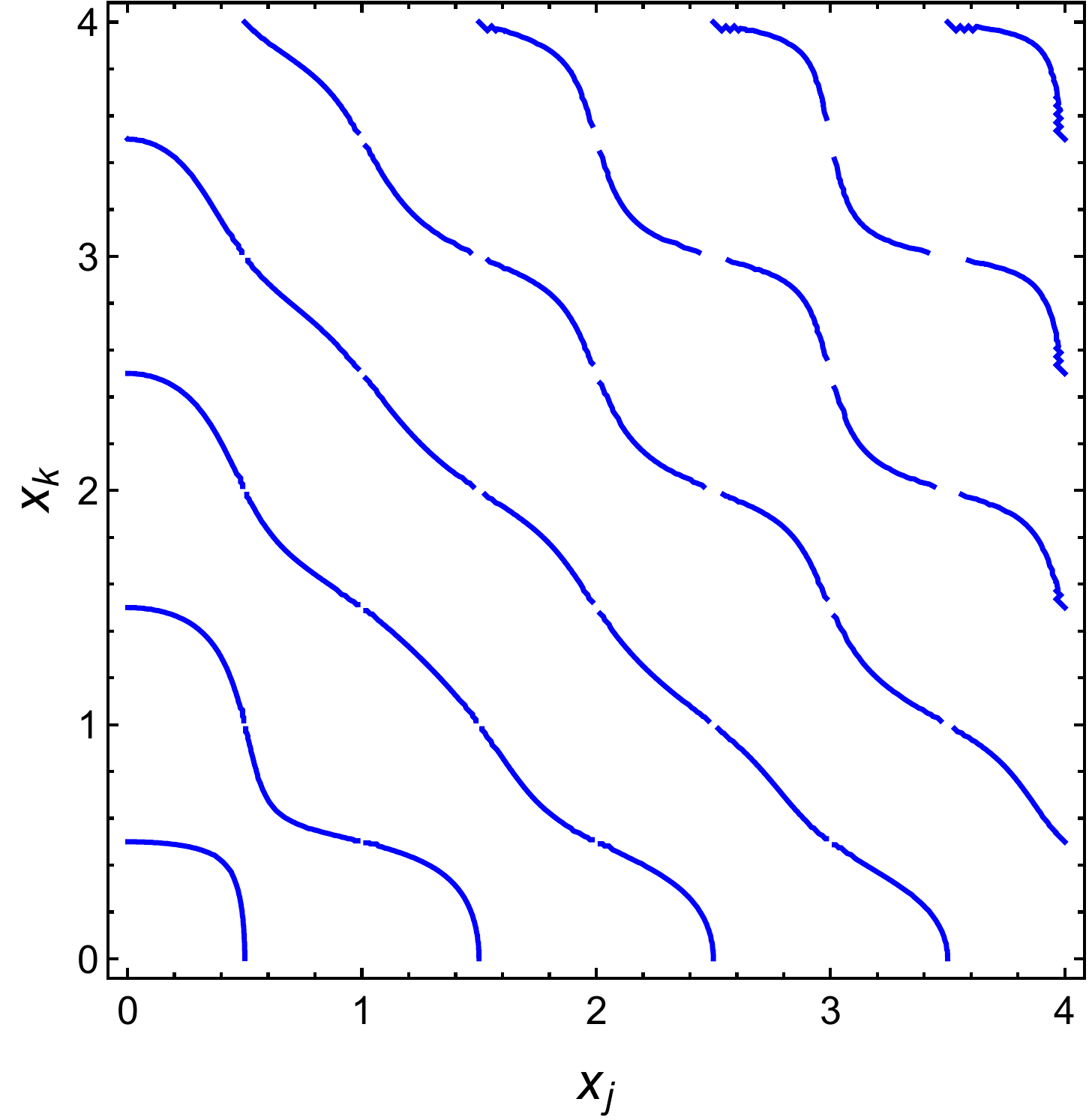}
    \caption{Solutions of the transcendental equation  (\ref{eq:root}) until $x_j\sim x_k\sim 3$ for $\delta_A=1$, for the thin-brane model. }
    \label{fig:rootEqA-example}
   \end{figure}

\begin{figure}
    \centering
    \includegraphics[scale=0.4]{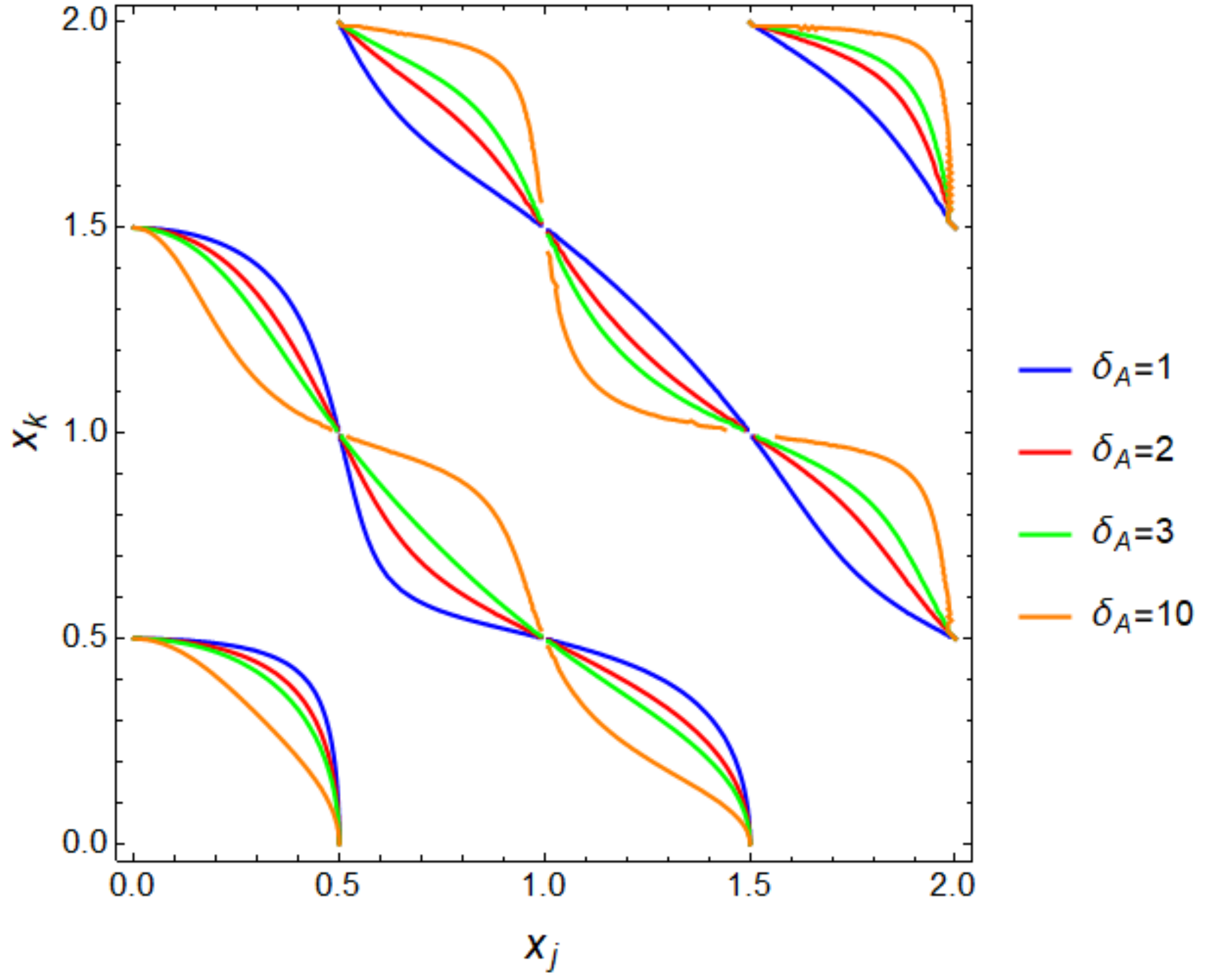}
    \caption{Solutions of the transcendental equation  (\ref{eq:root}) for different values of $\delta_A$, for the thin-brane model.}
    \label{fig:rootEqA}
\end{figure}

  Due to the presence of a BLKT the surface terms are no longer zero. The non-trivial solution ($\delta_A\neq 0$) for the Eq. (\ref{eq:vx4x5delta}) is
  \begin{eqnarray}
      v_0^{(j,k)}(x^4,x^5)=&A_{j,k}\Big[\cos (m_j x^4)\cos (m_k x^5)\nonumber\\
    &+\cos (m_k x^4)\cos (m_j x^5)\Big]\nonumber\\&+B_{j,k}\Big[\sin (m_j x^4)\sin (m_k x^5)\nonumber\\
    &+\sin (m_k x^4)\sin (m_j x^5)\Big]\,.
  \end{eqnarray}
 The solution above no longer satisfy the last BC in Eq. (\ref{eq:bcAmu}). Eq. (\ref{eq:v0}) is recovered if $A_{j,k}=-B_{j,k}$ and $\sin (m_k x_4)\sin (m_j x_5)$ is replaced by $-\sin (m_k x_4)\sin (m_j x_5)$. The coefficients $A_{j,k}$ and $B_{j,k}$ are found requiring the familiar conditions of continuity of the function and discontinuity of its derivative at $(0,0)$. Similar to the case of a delta-function in 1-D, we integrate the equation of motion (\ref{eq:vx4x5delta}) over $x_4$ and $x_5$, from $(0^-,0^-)$ to $(0^+,0^+)$. Performing a replacement of dummy variables we get
 \begin{eqnarray}\label{eq:vx4x5deltaInt}
  & \int_{0^-}^{0^+} \, dy\Big[\partial_4 v_0^{(j,k)}(x^4,y)|_{x^4=0^-}^{x^4=0^+}+\partial_5v_0^{(j,k)}(y,x^5)|_{x^5=0^-}^{x^5=0^+}\nonumber\\&- \partial_4 \overline{v}_0^{(j,k)}(x^4,y)|_{x^4=0^-}^{x^4=0^+}-\partial_5\overline{v}_0^{(j,k)}(y,x^5)|_{x^5=0^-}^{x^5=0^+}\Big]=\nonumber\\
    &- M_{j,k}^2\delta_AR^2v_0^{(j,k)}(0,0)\,,
  \end{eqnarray}
where $v_0^{(j,k)}$ is the wave-function for $x^4,x^5>0$ and $\overline{v}_0^{(j,k)}$ is the wave-function for $x^4,x^5<0$. Terms with crossed coordinates such as $\sim v_0^{(j,k)}(0^+,0^-)$ are zero. Using Eqs. (\ref{eq:vx4x5deltaInt}) and (\ref{mass}) we get the wave-function due to a two-dimensional delta-function source
 \begin{eqnarray}\label{eq:v0deltafunction}
      v_0^{(j,k)}(x^4,x^5)=&N_{j,k}\Big[\cos (m_j x^4)\cos (m_k x^5)\nonumber\\
    &+\cos (m_k x^4)\cos (m_j x^5)\nonumber\\&-\frac{\delta_A}{2}x_j x_k\Big (\sin (m_j x^4)\sin (m_k x^5)\nonumber\\
    &+\sin (m_k x^4)\sin (m_j x^5)\Big)\Big]\,,
  \end{eqnarray}
  where $m_j=x_j/R$, $m_k=x_k/R$ and $N_{j,k}$ is the normalization constant defined through Eq. (\ref{eq:normalizationcondit}), which gives
  \begin{eqnarray}
      N_{j,k}^{-2}=&\frac{\pi ^2R^2}{2}\biggr \{1 +\frac{\delta_A}{4\pi ^2}  \cos ^2(\pi  x_j)\Big[1+\cos ^2(\pi  x_k)\Big]\nonumber\\
    &+\frac{  \sin (2 \pi  x_k)}{2 \pi x_k}+\frac{1}{4}  \delta_A^2 x_j^2 x_k^2\nonumber\\&-\frac{ \delta_A}{2\pi}\Big[ x_k \cos ^2(\pi  x_j) \cot(\pi  x_k)+ x_j \cot (\pi  x_j) \cos ^2(\pi  x_k)\Big]\nonumber\\
    &-\frac{x_j x_k \sin(2 \pi  x_j) \sin (2 \pi  x_k)}{ \pi^2(x_j^2-x_k^2){}^2}\nonumber\\&+\frac{4 x_k^2 \cos ^2(\pi  x_j) \csc ^2(\pi  x_k)}{\pi^2(x_j^2-x_k^2){}^2}+\frac{4 x_j^2 \csc ^2(\pi  x_j) \cos ^2(\pi  x_k)}{\pi^2(x_j^2-x_k^2){}^2}\biggr\}\,.
  \end{eqnarray}
As in the 5-D case \cite{Rizzo:2018ntg}, the transcendental equation that determines the roots $x_j$ and $x_k$ is found requiring the Dirichlet BC $ v_0^{(j,k)}(\pi R,\pi R)=0$, whose solutions depend only upon $\delta_A$
\begin{equation}\label{eq:root}
    \cot(\pi  x_j)\cot(\pi  x_k)=\frac{\delta_A}{2}x_j x_k\,.
\end{equation}
Eq. (\ref{eq:root}) has an evident resemblance to the  root equation in 5-D ($\cot(\pi  x_n)=\delta_Ax_n/2$) \cite{Rizzo:2018ntg}. Since only one equation  (\ref{eq:root}) determines both roots $x_j$ and $x_k$, it is expected the existence of a continuous set of values $x_j$ and $x_k$ that satisfies Eq.  (\ref{eq:root}). The solutions of Eq.  (\ref{eq:root}) are shown in Fig. \ref{fig:rootEqA-example}, for $\delta_A=1$, while different values of $\delta_A$ are plotted in Fig. \ref{fig:rootEqA}. 

From Fig. \ref{fig:rootEqA-example} we see that there are $(2n+1)$ quantized masses for each curve  $n$, where we labeled $n$ as being each one of the dashed lines. Each mode is described by the segments in the dashed lines, \textit{i.e.}, at $n=0$ (first dashed line) there is one mode $M_{0,0}$, a massive zero-mode, the second dashed line ($n=1$) has three  quantized masses $M_{0,1}$, $M_{1,0}$ and $M_{1,1}$ (being the first two degenerate), and so on. The segments in the middle of each  dashed line are the levels corresponding to $M_{j,j}$ and since the curves are symmetric under reflection over the line $x_j=x_k$, the masses $M_{j,k}$ and $M_{k,j}$ are degenerate. These features are the usual behavior of quantum systems in two dimensions. Although there is a continuous set of values $(x_j, x_k)$  in each segment, the whole set represent only one  (mass) state, being narrow the range of each state. In Table \ref{tab:table} it is presented the masses $M_{j,k}$ for the first three curves of Fig. \ref{fig:rootEqA-example}. The masses correspondent to each KK level are either increased or decreased in an alternated pattern, when the parameter $\delta_A$ is increased, as seen in Fig. \ref{fig:rootEqA}. 
\begin{table}[]
    \centering
        \begin{tabular}{|c| c |}
        \hline
      $(j,k)$   &$ M_{j,k}R\,$\\\hline $(0,0)$&$0.5-0.6$ \\\hline $(1,1)$ & $0.9-1.1$\\ $(0,1)$&  $1.1-1.5$\\$(1,0)$& $1.1-1.5$\\\hline $(2,2)$&$1.8$ \\ $(0,2)$&$1.8-2.1$\\ $(1,2)$& $2.1-2.5$\\ $(2,0)$& $1.8-2.1$\\ $(2,1)$ &$2.1-2.5$\\\hline
         \end{tabular}
    \caption{Mass range $ M_{j,k}R=\sqrt{x_j^2+x_k^2}$ for the first three curves plotted in Fig. \ref{fig:rootEqA-example}. }
    \label{tab:table}
\end{table}

\subsection{BLKT at (0,0) and ($\pi R$, $\pi R$)}
\begin{figure*}
    \centering
    \includegraphics[scale=0.42]{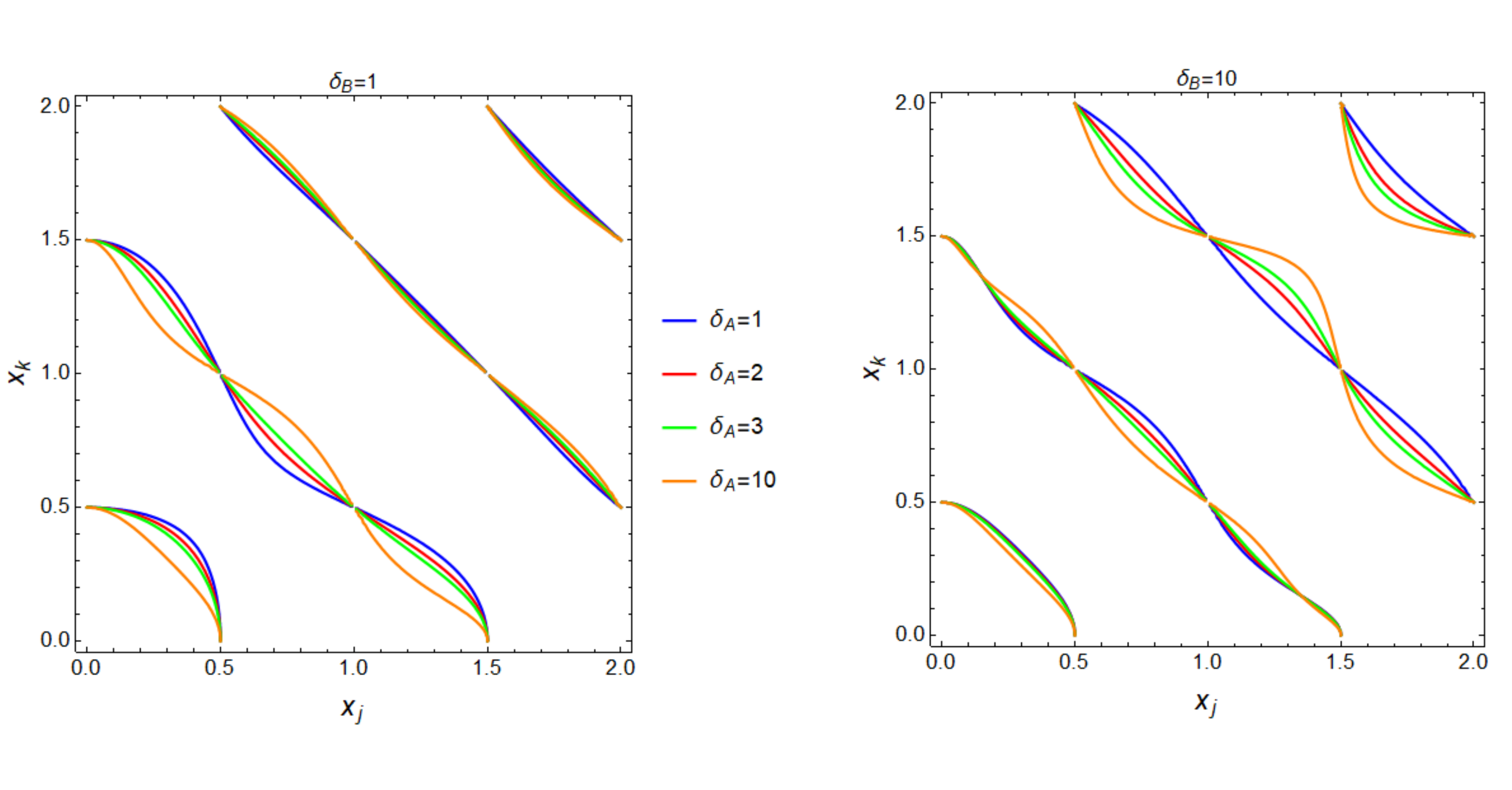}
    \caption{Solutions of the transcendental equation  (\ref{eq:rootAB}) for different values of $\delta_A$ and two values of $\delta_B$, for the thin-brane model.}
    \label{fig:rootEqAB}
\end{figure*}

\begin{figure}
    \centering
    \includegraphics[scale=0.45]{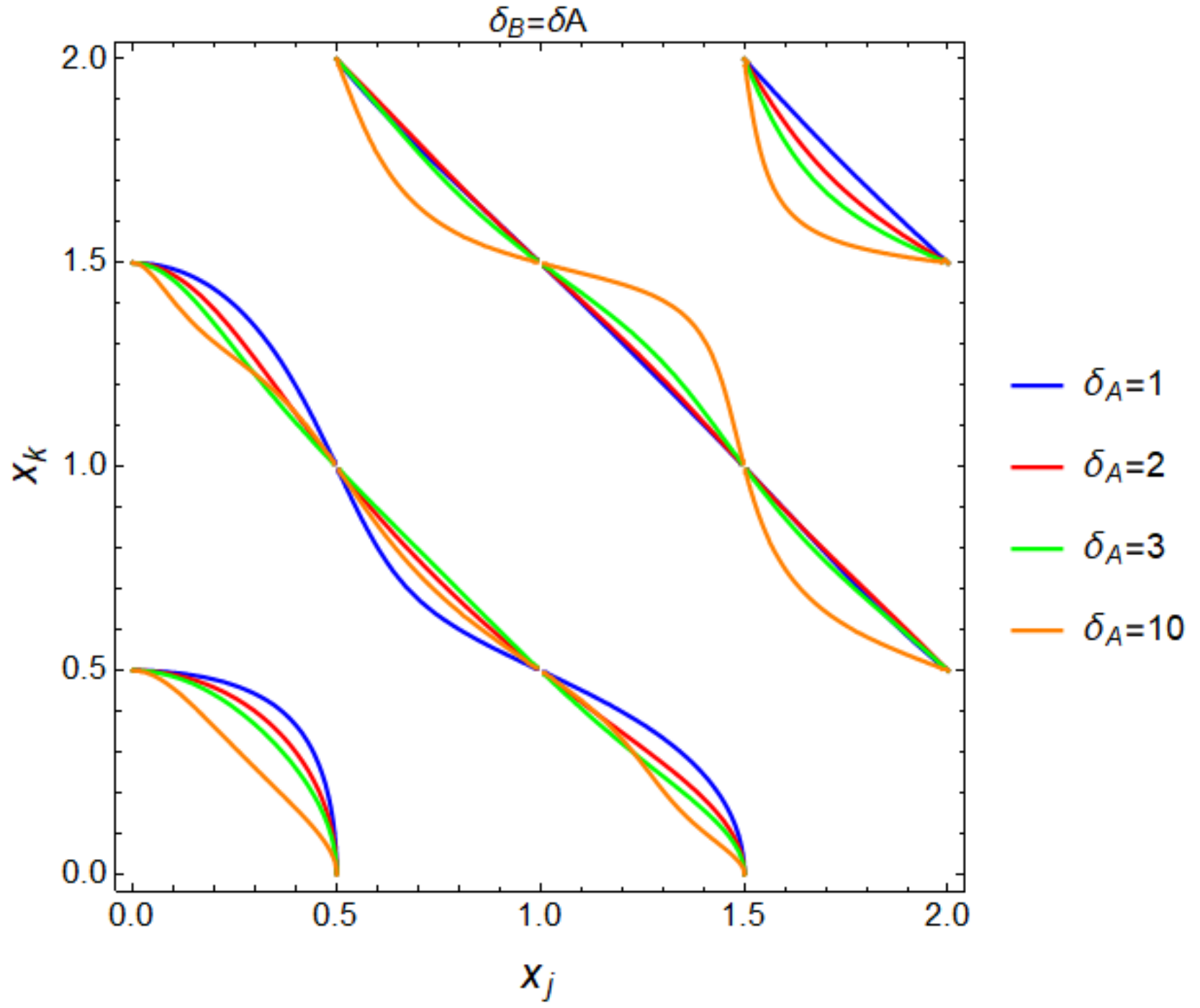}
    \caption{Solutions of the transcendental equation  (\ref{eq:rootAB}) for different values of $\delta_A$, when $\delta_A=\delta_B$, for the thin-brane model.}
    \label{fig:rootEqABequal}
\end{figure}

We consider now branes localized at $(0,0)$ and $(\pi R, \pi R)$ with BLKT on them. For the sake of completeness we add the following term in the Lagrangian
\begin{eqnarray}
\mathcal{L}_{BLKT}=&-\frac{1}{4}V_{\mu\nu}V^{\mu\nu}\cdot\Big[ \delta_A R^2\,\delta(x^4,x^5)\nonumber\\
    &+\delta_B R^2\, \delta(x^4-\pi R, x^5-\pi R)\Big]\,,
\end{eqnarray}
where $\delta_A$ is not necessarily equal to $\delta_B$. The equation of motion (\ref{eq:vx4x5delta}) is modified by an extra term proportional to $\delta_B$. The normalization factor has now the following terms
\begin{equation}
    Z_{(j,k)}=1+\delta_A R^2 v_0^{(j,k)}(0,0)+\delta_B R^2 v_0^{(j,k)}(\pi R,\pi R)\,.
\end{equation}
The wave-function is equal to Eq. (\ref{eq:v0deltafunction}) for $x^4,x^5\leq \pi R$, and the transcendental equation is found through the non-continuity of the derivative of the wave-function, whose expression is similar to  (\ref{eq:vx4x5deltaInt}). The quantized masses are therefore found  through the transcendental equation
\begin{equation}\label{eq:rootAB}
    \left(1+\frac{\delta_A \delta_B}{4}x_j^2x_k^2\right)\cot(x_j \pi)\cot(x_k \pi)=\frac{x_j x_k}{2}(\delta_A+\delta_B)\,,
\end{equation}
which is reduced to Eq. (\ref{eq:root}) when $\delta_B=0$. This root equation is also similar to the one in the 5-D case \cite{Carena:2002me}. The solutions of Eq. (\ref{eq:rootAB}) are depicted in Fig. \ref{fig:rootEqAB}. For lower (higher) values of $\delta_B$ the  larger (smaller) roots start having the same value, roughly independent of $\delta_A$. The case $\delta_A=\delta_B$ preserves KK-parity and the roots are presented in Fig. \ref{fig:rootEqABequal}. Their values are similar to the case  $\delta_B=0$.

\subsection{BLKT at $(0,\pi R)$}
Since the points $(0,\pi R)$ and $(\pi R,0)$ are identified it is sufficient to consider only one case. We will consider now the BLKT inside a brane localized at $(0,\pi R)$, whose Lagrangian is
\begin{equation}
\mathcal{L}_{BLKT}=-\frac{1}{4}V_{\mu\nu}V^{\mu\nu}\cdot \delta_A R^2\,\delta(x^4,x^5-\pi R)\,.
\end{equation}
Similar to the previous cases, the normalization constant becomes
\begin{equation}
    Z_{(j,k)}=1+\delta_A R^2 v_0^{(j,k)}(0,\pi R)\,.
\end{equation}
The solution for the equation of motion with a delta function source at $(0,\pi R)$, satisfying similar BC as Eq. (\ref{eq:vx4x5deltaInt}), is
 \begin{eqnarray}\label{eq:v0delta0piR}
     v_0^{(j,k)}(x^4,x^5)=N_{j,k}\Big[&\cos (m_j x^4+m_k x^5)+\cos(m_k x^4-m_j x^5)\nonumber\\
     &+\sin (m_j x^4+m_k x^5)\nonumber\\
    &+\sin(m_k x^4-m_j x^5)\Big]\,,
  \end{eqnarray}
  where
  \begin{eqnarray}
   N_{j,k}^{-2}&= 2\pi^2 R^2\biggr\{1+\frac{1}{\pi^2(x_j^2-x_k^2)}+\frac{\sin ^2(\pi  x_j) \sin (2 \pi  x_k)}{2\pi^2 x_j x_k}\nonumber\\
    &+\frac{ \cos (2 \pi  x_k)- \sin(2 \pi  x_j)}{\pi^2(x_j^2-x_k^2)}\nonumber\\
    &+\frac{2 \cos (\pi  x_k)}{\pi^2(x_j^2-x_k^2)}\Big[\sin (\pi  x_j) - \cos(\pi  x_j)\Big]\biggr \}   \,.
  \end{eqnarray}
     The dependence of $\delta_A$ appears in the transcendental equation, which is identical to Eq.  (\ref{eq:root}), thus having the same solutions for the pair of roots $(x_j,x_k)$.

\section{BLKT on a thick brane}\label{blkt-thick}
We consider now the effect of a BLKT on the thick brane, lying between $ \pi r < x^4,x^5 \leq \pi R$, with a width $\pi(R-r)\equiv \pi L$. The thin brane is obtained in the limit $L\rightarrow 0$. In 5-D, the difference between thin and thick branes leads to the suppressed coupling mechanism \cite{Landim:2019epv}, existing for branes with a finite thickness. In 6-D,  thin branes carry localized operators  on the conical singularities at the corners of the square. On the other hand, BLKT are spread inside the thick branes, thus for thin branes the surface terms (\ref{eq:surfaceterms}) are non-zero, while for a fat brane the operators are not localized at the conical singularities and there are two regions in the two-dimensional (ED) space, each one having (in principle) vanishing surface terms.    

The BLKT  with gauge fixing term is  \cite{Dvali:2000rx,Dvali:2001ae}
\begin{equation}\label{eq:lagrangianBLKT}
\mathcal{L}_{TB}=\left[-\frac{1}{4}V_{\mu\nu}V^{\mu\nu}-\frac{1}{2\xi}(\partial_\mu V^\mu)^2\right]\cdot \delta_A R^2\,\theta(x^4,x^5)\,,
\end{equation}
where the step function is non-zero only inside the thick brane, \textit{i.e.}
  \begin{eqnarray}\label{eq:theta}
      &\theta(x^4,x^5)=\alpha^2 \quad \text{for } \pi r < x^4,x^5 \leq \pi R\,, \qquad \nonumber\\
    &\theta(x^4,x^5)=0\quad \text{for } x^4,x^5<\pi r\,.
  \end{eqnarray}
  The equation of motion for the wave-function inside the thick brane is now
 \begin{equation}\label{eq:vx4x5alpha}
     \Big[\partial_4^2+\partial_5^2+( M_0^{(j,k)})^2+( M_0^{(j,k)})^2\delta_A\alpha^2R^2\Big]\overline{v}_0^{(j,k)}(x^4,x^5)=0\,.
  \end{equation}
  Similar to the 5-D case \cite{Landim:2019epv}, we may define an effective mass as $\overline{M}_0^{(j,k)}\equiv  M_0^{(j,k)}\sqrt{1+ \delta_A\alpha^2 R^2 }$, thus the presence of the step-function changes the mass term in the equation of motion for $V_\mu^{(j,k)}$ inside the thick brane.  It has the same structure of Eq. (\ref{eq:vx4x5}), but with the replacement $M_0^{(j,k)}\rightarrow \overline{M}_0^{(j,k)}$ \cite{Landim:2019epv}. Defining the effective mass parameters as
  \begin{equation}\label{massbarV}
    \overline{m}_j\equiv  m_j\sqrt{1+ \delta_A\alpha^2 R^2 }\,, \quad  \overline{m}_k\equiv  m_k\sqrt{1+ \delta_A\alpha^2 R^2 }\,,
\end{equation}
 the wave-function inside the thick brane   $\overline{v}_0^{(j,k)}(x^4,x^5)$  has also the same structure of Eq. (\ref{eq:v0}). The wave-function outside the thick brane is
  \begin{equation}\label{eq:v01}
     v_0^{(j,k)}(x^4,x^5)=A_1^{(j,k)}\left[\cos  (m_j x^4+m_k x^5)\pm\cos (m_k x^4-m_j x^5)\right]\,,
  \end{equation}
  while inside the thick brane the wave-function is
   \begin{equation}\label{eq:v02}
     \overline{v}_0^{(j,k)}(x^4,x^5)=A_2^{(j,k)}\left[\cos (\overline{m}_j x^4+\overline{m}_k x^5)\pm\cos (\overline{m}_k x^4-\overline{m}_j x^5)\right]\,,
  \end{equation}
  where $A_1^{(j,k)}$ and $A_2^{(j,k)}$ are coefficients to be determined. 
  
  Both wave-functions have this form in order to satisfy the BC (\ref{eq:bcAmu}). The mass parameters, however, should be either $\overline{m}_i=m_i=i/R $ ($+$) or  $\overline{m}_i=m_i=(i+1/2)/R$  ($-$),  for $i=j,k$,  in order to satisfy the same BC. This is only possible if $\delta_A\alpha=0$. Even if we assume that the fields no longer need to satisfy all previous BC, the situation remains the same by the following reason. The wave-function should be continuous at $(y,\pi r)$ and $(\pi r, y)$, as well as its derivative with respect to both ED coordinates $x^4$ and $x^5$. Thus $\overline{v}_0^{(j,k)}(\pi r,y)=v_0^{(j,k)}(\pi r,y)$ and $\partial_4 \overline{v}_0^{(j,k)}|_{(x^4,x^5)=(\pi r,y)}=\partial_4v_0^{(j,k)}|_{(x^4,x^5)=(\pi r,y)}$ (the continuity conditions at $(y, \pi r)$ give exactly the same expressions). These conditions can be satisfied at a point $y_c$ on the boundary, but it is not possible to match the functions all along the boundary, being the only possibility $\overline{m}_i=m_i$. Therefore the only viable solution is $\delta_A\alpha=0$, which leads to a thin brane.

\section{ Conclusions}\label{sec:discussion}

In this paper we have investigated the implications of BLKT on thin and thick branes, for a model of two ED compactified on the chiral square, when a vector field is present in the bulk. For thin branes the presence of BLKT gives mass to all modes of the KK tower of states, being the masses dependent upon the compactification radius and the BLKT parameter. The roots are roughly the same for branes at different positions, \textit{i.e.}, they have similar values for branes localized at $(0,0)$, $(0,\pi R)\sim(\pi R,0)$ and $(\pi R,\pi R)$. The transcendental equations and other relations resemble the 5-D case. The model presents the usual behavior of quantum systems in two dimensions, \textit{i.e.}, there are $(2n+1)$ quantized masses for each curve  $n$, and each mode is described by the segments in the dashed lines: one massive zero-mode  $M_{0,0}$ at $n=0$, three  quantized masses $M_{0,1}$, $M_{1,0}$  and $M_{1,1}$ at $n=1$ (being  $M_{0,1}$ and $M_{1,0}$ degenerate), etc. 

The BLKT on thick branes, on the other hand, does not provide a non-trivial result $\delta_A\alpha\neq 0$ due to the BC. This scenario is similar to the refraction of a wave-function by a two-dimensional step function. Suppose   an incident wave-function $e^{i(k_x x+ k_y y) }$, a refracted wave  $e^{i(q_x x+ q_y y) }$ and a step-function different from zero for $x>0$. The BC implies that $k_y=q_y$ and the analogue situation in our problem is therefore $\Bar{m}_i=m_i$. Hence the mechanism of suppressed coupling in 5-D \cite{Landim:2019epv} cannot be applied in 6-D. 

The results presented here works for different fields in the bulk and can be used in several further proposals, as for instance, in a model of ED with the dark photon as mediator. This model was done in 5-D \cite{Rizzo:2018joy,Rizzo:2018ntg}, but an extension might be able with two ED as well, or even its inclusion in the UED model. In both cases, the BLKT breaks the extra $U(1)_D$ gauge symmetry via BC without adding an extra Higgs-like field, avoiding, in turn,  constraints on the Higgs-portal coupling. Potential signatures for such massive spin-1 KK particles depend upon the specific model considered but it usually includes missing energy searches, which might constrain the two parameters in this model.

\begin{acknowledgements}
The author would like to thank Gia Dvali for clarifications and Thomas Rizzo for comments. This work was supported by CAPES under the process 88881.162206/2017-01 and Alexander von Humboldt Foundation. \end{acknowledgements}

\bibliographystyle{unsrt}
\bibliography{references}\end{document}